\documentclass[sigconf,authorversion,screen,nonacm]{acmart}

\usepackage{booktabs}

\acmDOI{xx.xxx/xxx_x}

\acmISBN{978-1-4503-9517-5/23/03}

\acmConference[SAC'23]{ACM SAC Conference}{March 27 –April 2, 2023}{Tallinn, Estonia}
\acmYear{2023}
\copyrightyear{2023}

\acmArticle{4}
\acmPrice{15.00}

\usepackage[bottom]{footmisc}

\begin{document}
\title{Detecting and Preventing Credential Misuse in OTP-Based Two and Half Factor Authentication Toward Centralized Services Utilizing Blockchain-Based Identity Management}

\renewcommand{\shorttitle}{Detecting and Preventing Credential Misuse in OTP-Based Two Factor Authentication Toward Centralized Services}

\author{Jozef Drga}
\affiliation{
  \institution{Brno University of Technology, Faculty of Information Technology}
      \country{Czech Republic}
}
\email{idrga@fit.vutbr.cz}

\author{Ivan Homoliak}
\affiliation{
  \institution{Brno University of Technology, Faculty of Information Technology}
    \country{Czech Republic}
}
\email{ihomoliak@fit.vutbr.cz}

\author{Juraj Van\v{c}o}
\affiliation{
  \institution{Tomas Bata University in Zl\'in, Faculty of Applied Informatics}
  \country{Czech Republic}
}
\email{vanco@utb.cz}

\author{Martin Pere\v{s}\'ini}
\affiliation{
  \institution{Brno University of Technology, Faculty of Information Technology}	
  \country{Czech Republic}
}
\email{iperesini@fit.vutbr.cz}

\author{Petr Han\'a\v{c}ek}
\affiliation{
	\institution{Brno University of Technology, Faculty of Information Technology}	
	\country{Czech Republic}
}
\email{hanacek@fit.vutbr.cz}

\renewcommand{\shortauthors}{Drga et al.}

\begin{abstract}
This work focuses on the problem of detection and prevention of stolen and misused secrets (such as private keys) for authentication toward centralized services. 
We propose a solution for such a problem based on the blockchain-based two-factor authentication scheme SmartOTPs, which we modify for our purposes and utilize in the setting of two and half-factor authentication against a centralized service provider.
Our proposed solution consists of four entities that interact together to ensure authentication: (1) the user, (2) the authenticator, (3) the service provider, and (4) the smart contract.
Out of two and a half factors of our solution, the first factor stands for the private key, and the second and a half factor stands for one-time passwords (OTPs) and their precursors, where OTPs are obtained from the precursors (a.k.a., pre-images) by cryptographically secure hashing.
We describe the protocol for bootstrapping our approach as well as the authentication procedure.
We make the security analysis of our solution, where on top of the main attacker model that steals secrets from the client, we analyze man-in-the-middle attacks and malware tampering with the client.
In the case of stolen credentials, we show that our solution enables the user to immediately detect the attack occurrence and proceed to re-initialization with fresh credentials.
\end{abstract}

\keywords{Blockchain-based identity management, verifiable credentials, tampering with the client, two factor authentication.}

\maketitle

\section{Introduction}
In the 21th century, the question of securing personal data becomes a part of our daily life.
Users have many accounts at various Internet services, so they have to keep passwords for all of them.
Most of such passwords are remembered by users.
To ease the remembering, passwords are often weak and re-used with with different service providers.
These facts simplify dictionary attacks.
A potential solution to the password problems might be resolved by third party authentication schemes that enable to delegate authentication to large service providers such as Facebook, Google, Amazon, etc.
The well-known solution for third party authentication is Open Authorization (OAuth)~\cite{OAuth}.
However, even OAuth protocol v2.0 is vulnerable to phishing attacks~\cite{oauth-attack3}, click-jacking~\cite{oauth-attack1}, and redirect URL attacks~\cite{oauth-attack2}.
An alternative to password-based authentication is public key cryptography (a.k.a., asymmetric cryptography).
However, malware was developed to steal private keys as well~\cite{b0}.

Securing of the authentication process is a key problem for protection of the user session and her private data.
Service providers pay significant amount of money to provide a high security to their services.

In this work, we prose the authentication approach that provides two and half factor authentication using blockchain to prevent unauthorized access using private key and one time passwords (OTPs).
We assume that private keys of users are bound to their identities through blockchain-based identity management schema through its so called \textit{verifiable credentials}.

\subsubsection*{\textbf{Contributions}}
Our contributions are as follows:
\begin{enumerate}
	\item We propose a novel approach for authentication, providing two and half factor authentication based on public key cryptography and modified version of one time passwords (OTPs).
	Our approach enables to provably and immediately detect misuse of user credentials.
	\item We perform security analysis of the proposed approach and demonstrate its security under assumed attacker model.
\end{enumerate}

\subsubsection*{\textbf{Organization}}
The paper is organized as follows:
In \autoref{sec:background}, we present the elementary principles of blockchain, DLT, Merkle tree, identity management, and SmartOTPs that our work is inspired from.
We provide the problem definition in \autoref{sec:problem}, where we moreover present our motivation.
Then, in \autoref{sec:proposed-approach}, we describe the proposed approach of the second and half factor authentication.
We present the proof-of-concept implementation in \autoref{sec:implementation}.
In \autoref{sec:security} we perform security analysis of our approach and \autoref{sec:discussion} discusses costs implications and performance implication of the proposed solution. 
In \autoref{sec:related}, we summarize related work and 
\autoref{sec:conclusion} concludes our paper.

\section{Background}
\label{sec:background}
\subsection{Blockchain}

Blockchain is a data structure represented by a distributed ledger that consists from transactions in the blocks.
The blocks are ordered and maintained by mutually untrusted nodes, which use the consensus protocol to agree on the order of the blocks.
Blockchain uses cryptographic hash to maintain data integrity and its immutability.
A transaction might contain orders transferring cryptotokens or pieces of application code (smart contracts) that is executed in decentralized fashion.

\subsection{Blockchain-Based Identity Management}

\begin{figure}[t]
	\centering
	\includegraphics[width=\linewidth]{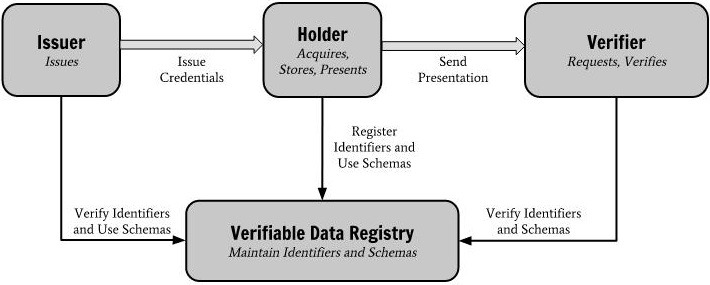}
	\caption{The model of verifiable credentials. Verifiable data registry is represented by an identity-oriented blockchain in a decentralized identity.}
	\label{fig:w3c-verifiable-cred}
	\vspace{-0.5cm}
\end{figure}

The identity and access management (IAM) defines and manages the roles and access privileges of individual users and the rules in which users are granted or denied those privileges~\cite{b5}.
Credentials are used for authentication in IAM systems.
Identity management based on blockchain uses special type of credentials, called verifiable credentials.
Verifiable credentials are the sort of electronic replacement of the paper credentials, such as the personal ID, driver license, verifiable diploma, etc.
The workflow of using verifiable credentials in blockchain-based IAM is depicted in \autoref{fig:w3c-verifiable-cred}, where the issuer (a.k.a., identity provider) issues verifiable credentials for the holders (i.e., users), who present them at any verifier (i.e., service provider).
All of these entities have access to the shared data registry represented by the dedicated \textit{identity-oriented blockchain},\footnote{A well known example of such a blockchain is Hyperledger Indy.} which contains only identifiers of holders, denoted as decentralized identifiers -- DIDs.\footnote{Note that DIDs themselves do not contain any private information about the user or their real-world identities.}
DIDs meet some interesting properties: they are globally unique, highly available, verifiable, and can be created by anyone upon request (i.e., censorship resistance).
A DID usually consists of the schema and the user's address (i.e. identifier) on that schema.
The schema represents the identifier of a blockchain plus the identifier of its particular chain (such as the main chain, the test chain, etc).

\subsubsection*{\textbf{Types of Identity Management}}
There are four types of identity management systems~\cite{bidm-categorization-1,dib-decentralized-identity,survey-identity-blockchain}, which we describe in the following.

\paragraph{a)\normalfont \textbf{Independent Identity Management}}
In this concept, each service provider has its own methods to declare identity of the user.
Knowledge of user identities among service providers is independent, for example, the identity could be declared with a combination of a username and password specific for a service provider.

\paragraph{b)\normalfont \textbf{Centralized Identity Management}}
The concept of centralized identity management is based on the existence of a trusted domain. All service providers in the trusted domain share the user's identity information.
The trusted domain is serviced with a single identity provider.

\paragraph{c)\normalfont \textbf{ Federalized Identity Management}}
In this concept multiple identity providers in one trusted domain are in a federation, so one provider is able to recognize an identity from another identity provider.

\paragraph{d)\normalfont \textbf{ Self-Sovereign Identity Management}}
Sovrin foundation describes self-sovereign identity~\cite{b6} as an ''\textit{Internet for identity where no one owns it, everyone can use it, anyone can improve it}''.
In other words, this concept is based on principle that users are able to control their own verifiable credentials.
Users can provide their verifiable credentials to third parties they wish to authenticate themselves to.
Users have a full control over their verifiable credentials.

\subsection{Creation and Utilization of Verifiable Credentials}
\label{sec:vc-generation}
Life-cycle of verifiable credentials has two phases: (1) creation of the verifiable credentials and (2) their utilization in authentication.
Identity providers serve as the issuers of verifiable credentials, which is described in the following:

\begin{enumerate}
	\item The user creates her DID (referring to her identity in a particular blockchain) in the identity-oriented blockchain (i.e., verifiable data registry in \autoref{fig:w3c-verifiable-cred}).
	\item The user asks issuer to verify and sign new verifiable credentials, while providing her public key, DID and identity confirming documents such as passport, ID card, etc.
	\item Issuer verifies the identity of the user based on the provided documents.
	\item Issuer signs requested credentials.
	Afterward, the issuer sends credentials to the user.
	Note that to granularize privacy of information in verifiable credentials, the issuer might (upon user request) create various tuples of verifiable credentials (that might be subsets of each other or slightly overlap, etc).
	\item User verifies received credentials with their signature and stores them locally in her credential repository.
\end{enumerate}

\noindent
Next, we describe the process of using the verifiable credentials in a scenario where the user needs to be age-verified by a service provider (e.g., e-shop with tobacco):

\begin{enumerate}
	\item The user navigates her browser to the website of the e-shop.
	\item E-shop requests proof-of-age from the user's browser.
	\item Browser displays this request to the user and using the local credentials repository, it shows her all verifiable credentials that meet the proof-of-age requirement.
	\item The user selects verifiable credentials and the browser sends them to the e-shop.
	\item E-shop verifies the age in the credentials and verifies their validity (e.g., whether they were not revoked) and signature of the identity provider who issued them.
	\item Upon success, the user  is redirected to the age-verified version of the e-shop.
\end{enumerate}

\subsection{Merkle Tree}
\begin{figure}[t]
	\centering
	\includegraphics[width=\linewidth]{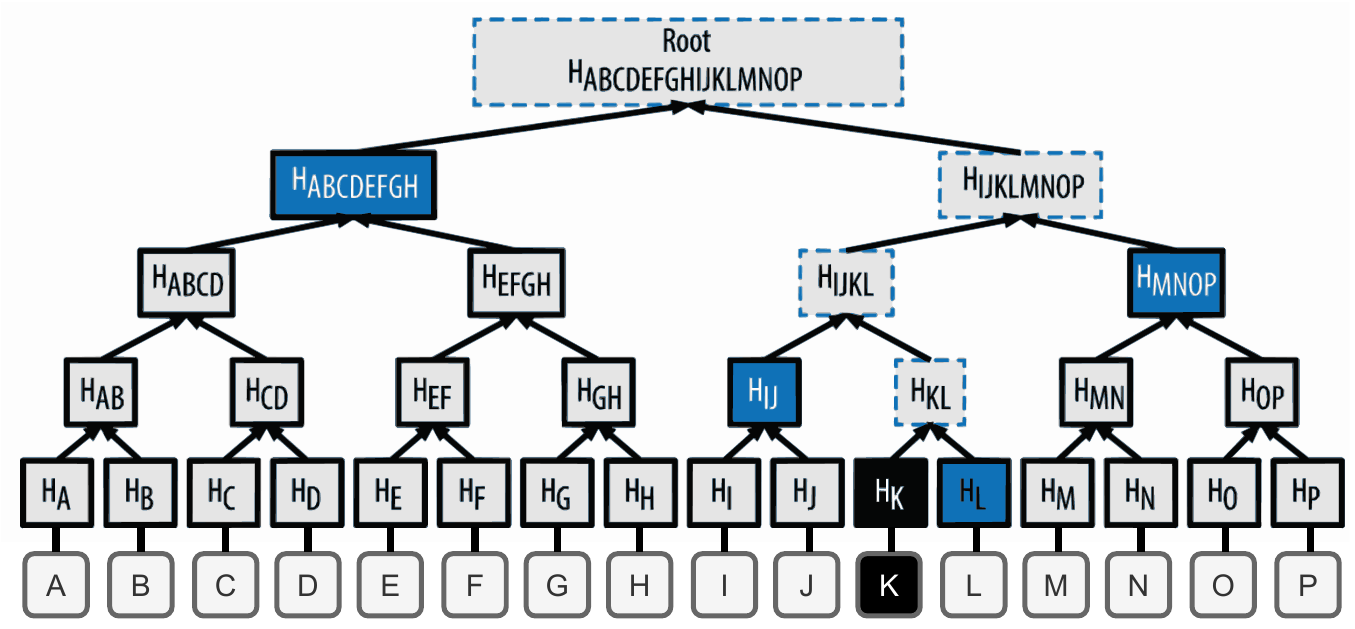}
	\caption{Example of the Merkle tree.}
	\label{fig:mt}
	\vspace{-0.5cm}
\end{figure}

Merkle tree is an integrity preserving data structure that uses cryptographic hash function to hierarchically aggregate the data.
In Merkle tree, each leaf node contains hash of the data, while each non-leaf node contains hash of its concatenated child nodes.
In this way, the root hash contains the integrity information about the whole tree and its data.
\autoref{fig:mt} depicts an example of the full binary Merkle tree with 16 data nodes and Merkle proof for data node K, which is depicted by blue background.

\subsection{SmartOTPs as the Second Factor of Authentication}
SmartOTPs~\cite{b13} is the approach proposing one time passwords (OTP) for blockchains with smart contract platforms; and thus they enable to enrich public key-based authentication by the second factor with interesting usability properties.
The two-factor authentication is performed in two stages of interaction with the blockchain, requiring two transactions -- one for each factor (i.e., public key and OTP).

In SmartOTPs, the OTPs are aggregated by the Merkle tree and hash chains, whereby for each authentication, only a short OTP (e.g., 16B-long) is transferred from the authenticator to the client.

Our work is inspired by the base version of SmartOTPs (not assuming the hash chains).
We denote  $ OTP_i^{'}$ as a precursor of $OTP_i$,  where $OTP_i$ is created as the hash of its precursor and represents a leaf of the Merkle tree in our approach.
Such a modification  allows us to store the leaves of the tree on a different device than OTP precursors, and thus utilize them independently.

\section{Problem Definition and The Attacker Model}
\label{sec:problem}
In this work, we focus on user authentication method against centralized services using blockchain-based identity management.
The solution we seek for should provide higher degree of security residing in the blockchain-based smart contract verifiability.
We want the user to be able of immediately figuring out whether her credentials-binded secret (i.e., private key) was stolen and utilized for the authentication at some centralized service.
Traditionally, this is a non-trivial problem since the user does not know whether her credentials were stolen and misused, unless a centralized service has a notification for each an every login attempt, which is not a common practice.

\subsubsection*{\textbf{Attacker Model}}
We consider the attacker model in which the attacker can steal all the user secrets stored in her client interface and attempt to misuse them for authentication.

Then, we assume that the user is communicating with the service provider through encrypted TLS sessions.
Also, we assume that the user is able to verify the service provider's identity and public key from her DID document and X.509 infrastructure; therefore, the user starts communication with the service provider only after successful verification.
Finally, we assume that the attacker is unable to break used cryptographic primitives nor blockchain platform.
 
\section {Proposed Approach}
\label{sec:proposed-approach}
We propose a new authentication approach based on Smart-OTPs~\cite{b13}, which enables two and half factor authentication (1 basing on public key cryptography and 1.5 basing on our modification of SmartOTPs).
Our approach utilize a blockchain with a Turing-complete smart contract capability. Alike SmartOTPs, we assume a usage of authenticator.
Authenticator can be an application running on smart phone or similar device.

\begin{figure}[t]
	\centering
	\includegraphics[width=0.75\linewidth]{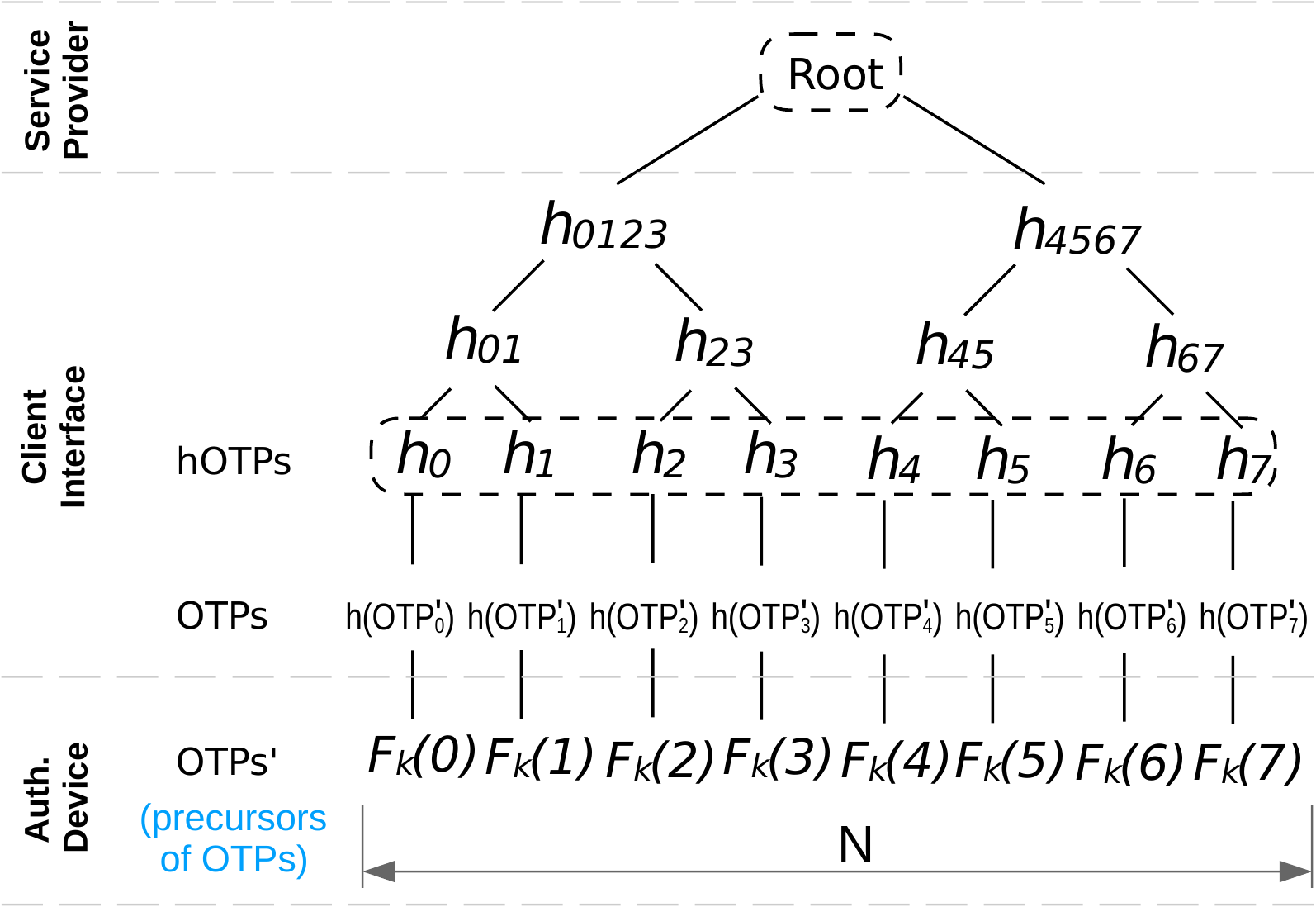}
	\caption{The scheme of OTP-related data distribution among  the parties of our approach. $F_k(.)$ represents a pseudo-random function with the secret seed $k$.}
	\label{fig:proposed}
	\vspace{-0.5cm}
\end{figure}

\medskip
We consider four entities in our solution: \textit{authenticator, the user (i.e., at the client interface), service provider, and the blockchain}.
Also, we assume that each service provider has her own smart contract deployed for our solution.
The first factor stands for a traditional public-key authentication.
The second one and half factor is based on SmartOTPs, and it is generated at the authenticator.
The authenticator generates a secret seed $k$.
From $k$, a list of $N$ OTPs is derived together with their precursors, denoted as ${OTPs'}$.
OTPs from the last iteration layer of SmartOTPs (i.e., \textit{OTPs} in \autoref{fig:proposed}) represent the second factor of authentication, while revealing the precursors of OTPs stands for another ``half'' factor of this process.
The situation is depicted in \autoref{fig:proposed}, where we can see how are data related to OTPs distributed among the components of our approach.
In detail, precursors OTPs' are stored at the authenticator, while OTPs (and all nodes of Merkle tree that can be derived from them) are stored at the user side -- in the client interface (e.g. user browser).
The service provider initially stores only the root hash of the Merkle tree aggregating OTPs.

\subsection{Bootstraping Phase}\label{sec:bootstrapping}

The protocol has a bootstrapping phase to establish binding among authenticator, client, and the service provider -- each of them storing only minimal information to execute authentication.
The communicating entities in the bootstrapping stage trust each other (i.e., we assume a secure bootstrapping, which is a common assumption).
The service provider deploys her own smart contract on the blockchain.
Then our protocol requires the registration of authenticator at the service provider.

\begin{figure}[t]
	\centering
	\includegraphics[width=\linewidth]{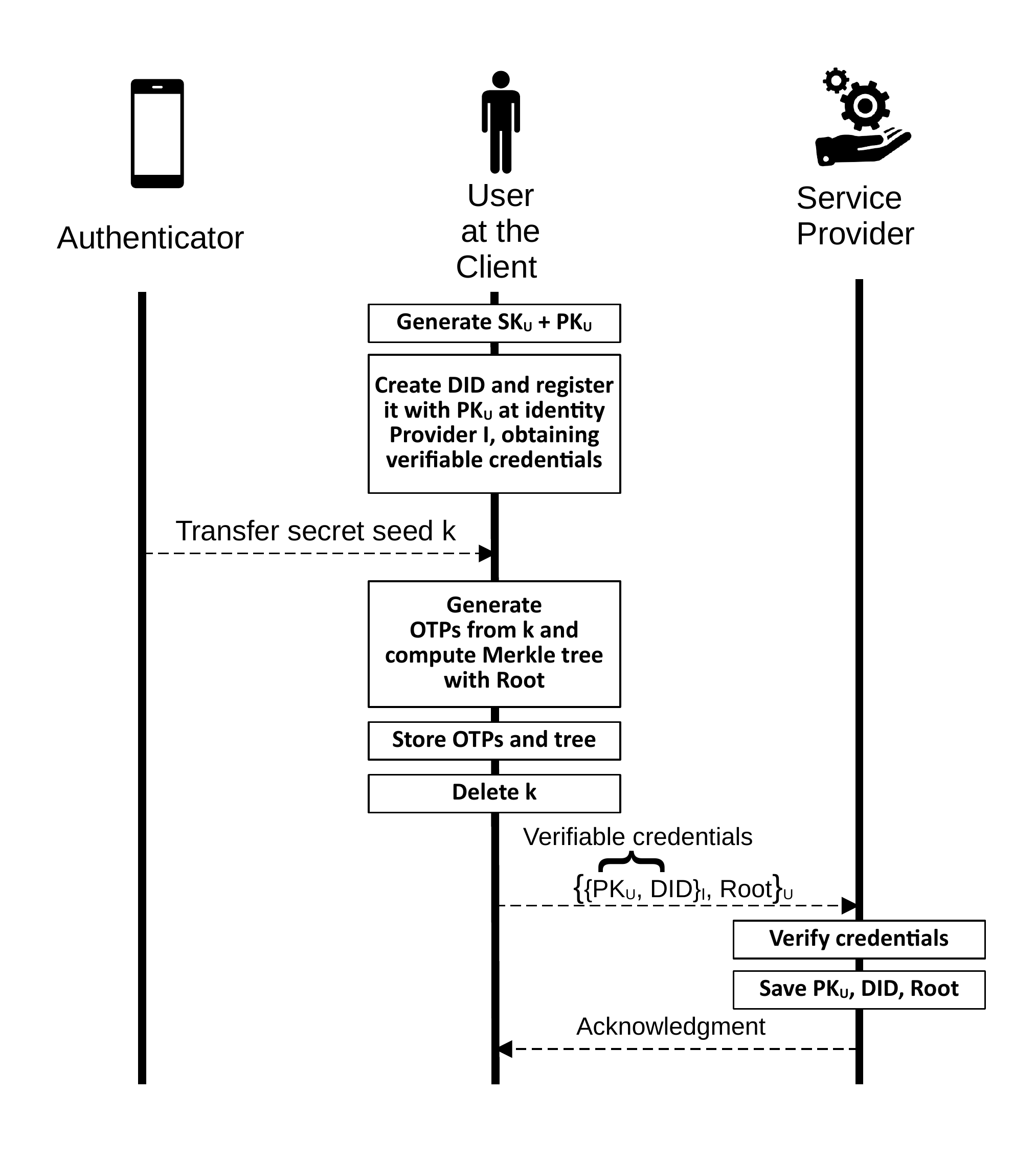}
	\vspace*{-1cm}
	\caption{Bootstaping phase of the proposed approach.}
	\label{fig:bp}
\end{figure}

In detail, the bootstrapping phase is described in \autoref{fig:bp} and contain several steps, which are as follows:
\begin{enumerate}
	\item The user generates the pair of public and private key at her client interface.

	\item The user creates her DID at identity-oriented blockchain and then registers it together with her public key at identity provider I, obtaining verifiable credentials (as described in Section~\ref{sec:vc-generation}).

	\item Authenticator generates the secret seed $k$.

	\item The user transfers $k$ to the client interface by an air-gapped transfer -- i.e., transcription of mnemonic words or scanning of the QR code (as proposed in SmartOTPs).

	\item $N$ precursors $OTPs^{'}$ are generated from $k$ by a pseudo\--random\footnote{$F_k(i)$ might be implemented by a cryptographically secure hash function $h(.)$ as $F_k(i) = h(k ~||~ i)$} function $F_k(i)$ (for $i \in 1,\ldots,N$) and then all $OTPs$ are computed from the precursors by cryptographically secure hashing (see Figure~\ref{fig:proposed}).
	The Merkle tree aggregating all $OTPs$ and its root hash is computed at the client.

	\item All nodes of the Merkle tree are stored at the client. This enables faster computation of Merkle proofs that are just fetched from the local storage of the client as opposed to the option where client were to store only OTPs and would compute the proofs on-the-fly.

	\item Client deletes the secret seed k and the precursors $OTPs'$.

	\item The client sends securely her verifiable credentials ( containing public key and her DID) together with the root hash of the Merkle tree to service provider within a signed TLS-encrypted message.

	\item Using a public key of the known identity provider $I$, service provider verifies the identity of the user from her verifiable credentials (containing signed DID and $PK_U$).

	\item Service provider saves the user's public key, DID, and the root hash of the Merkle tree.

	\item The service provider sends the acknowledgment of the registration to the client.
\end{enumerate}

\subsection{Operational Phase}
\begin{figure}[t]
	\centering
	\includegraphics[width=0.86\linewidth]{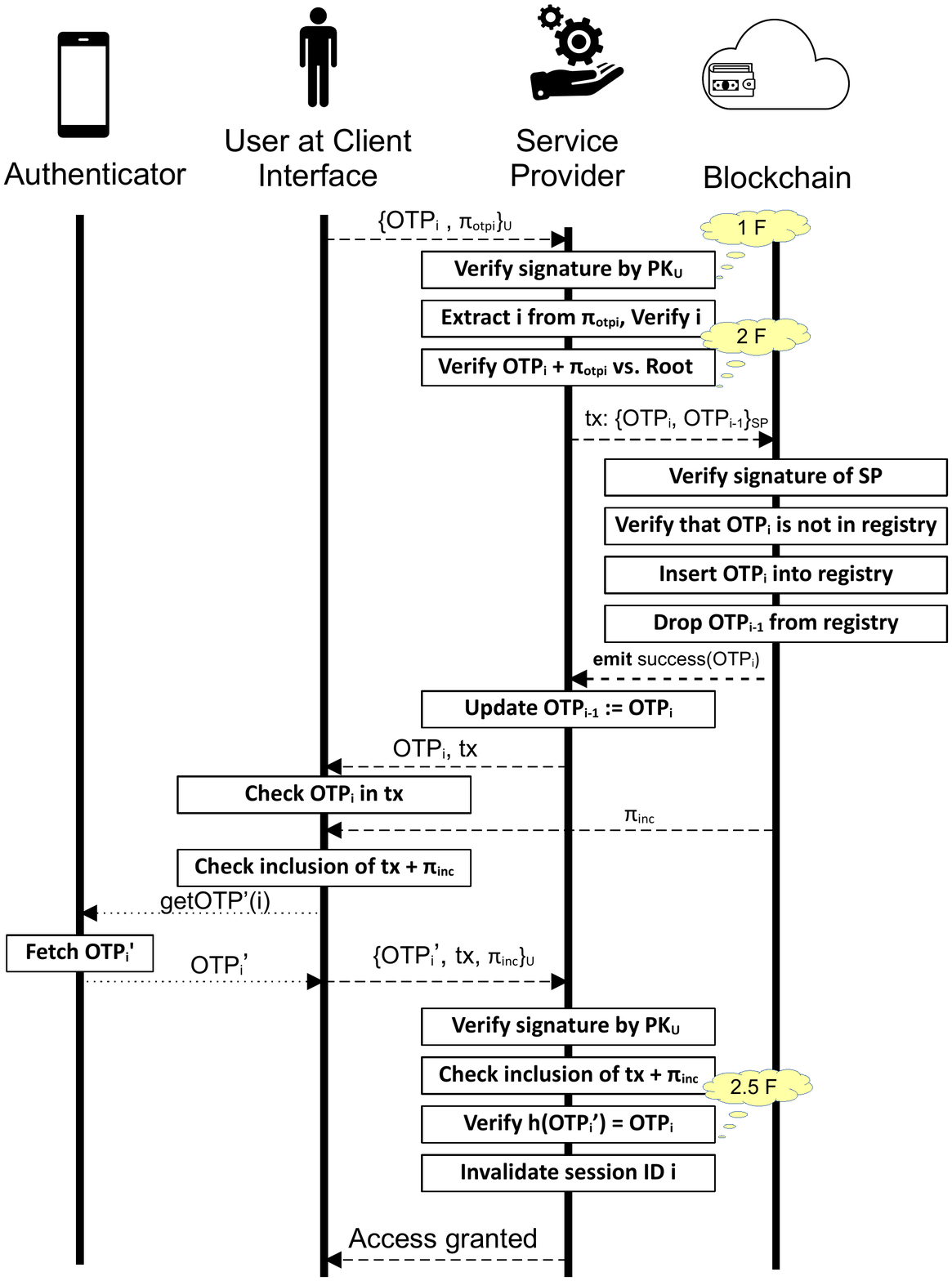}
	\vspace*{-0.8cm}
	\caption{Operational phase of the proposed approach, representing the authentication procedure.}
	\label{fig:op}
\end{figure}

We assume that the bootstrapping stage that registers the user  at the service provider was successfully finished.
Now, we describe the protocol for authentication of the user at the service provider.
The protocol is depicted in \autoref{fig:op}.

\begin{enumerate}
	\item Client sends the TLS-encrypted signed message containing $OTP_i$ and its Merkle proof $\pi_{OTP_i}$ to the service provider, where $i$ is the increment only counter of successful user sessions and it matches with the OTP ID in the Merkle tree.
	Both of these items are extracted from the client interface of the user.
	Also, $SK_U$ is fetched from the client interface and used to make a signature.
	$OTP_i$ together with a signature form two factors of the authentication.

	\item Service provider verifies the signature of the user against $PK_U$ that was extracted from verifiable credentials during bootstrapping.

	\item Service provider verifies the membership of the $OTP_i$ in the Merkle tree.
	In detail, using $OTP_i $ and its proof $\pi_{OTP_i}$, service provider computes the root hash of the tree and compares it with the stored root hash.

	\item Service provider verifies whether the previous  session ID stored by her has been completely authenticated (thus invalidated) and whether it is equal to i-1.
	If so, the service provider increments the session ID associated with the user (to value i), indicating the current session that is above to be opened.
	Otherwise, service provider sends an alert about the attempted misuse of the session back to the user.
	Note that the user can verify anytime on her own that the session was (unsuccessfully) misused by checking the registry of smart contract on the presence of $OTP_{i}$.

	\item Service provider publishes the $OTP_i$ on the blockchain via her smart contract.
	Moreover, service provider fetches the previously submitted OTP of the user from user data record and appends it to the transaction.
	Note that every service provider has its own smart contract deployed and items in the registry contain only (recent) OTPs of all users that the service providers handles.
	Therefore, these OTPs and insertion of them to registry cannot be linked to the particular user identity (or $PK_U$).
	The only information that is visible to public is that someone tries to authenticate at a particular service provider, while the public is also aware about the number of users the service provider handles\footnote{More precisely, it is the number of registered users who already have done at least one authentication attempt.} -- this number is equal to the number of items in the registry of the last used OTPs.

	\item Smart contract verifies whether $OTP_i$  is in global registry of last used OTPs.
	If so, the process is aborted and the smart contract returns attempted credentials misuse event to the service provider, who forwards it to the client.
	Alternatively, the user might also listen to the events of smart contract and receive this information.
	In the successful case, the smart contract inserts $OTP_i$ to its registry, removes the last inserted OTP of the user (i.e., $OTP_{i-1}$), which helps to delete unnecessary storage from the blockchain.
	Finally, the smart contract returns success together with transaction ID of the current call.

	\item Service provider updates $OTP_{i-1} := OTP_{i}$ in its user data record, which is always appended to the next transaction publishing $OTP_{i}$.

	\item Service provider returns the $OTP_i$ and transaction to the client.

	\item The client checks whether returned transaction contains correct data (related to $OTP_i$).
	Then, the client fetches the inclusion proof $\pi_{inc}$ of the transaction from the blockchain and using her
	light client checks whether the transaction was indeed included in the blockchain.

	\item User enters $i$ to the authenticator that will display the OTP precursor (i.e., $OTP_i^{'}$) encoded as mnemonic string or QR code -- preserving the air-gapped settings from Smart-OTPs.

	\item The user transfers $OTP^{'}_i$ from the authenticator to the client interface by transcription of mnemonic string or scanning of QR code by client machine's camera.

	\item The user sends the signed TLS-encrypted message containing transaction, its inclusion proof, and $OTP_i^{'}$ to the service provider.

	\item Service provider verifies the signature of the user.

	\item Service provider verifies whether the Session ID $i$ extracted from the corresponding proof of $OTP_i$ is an initiated valid session.
	If not, it aborts the process and returns failure.

	\item Service provider verifies the precursor of $OTP_i$ by checking equality $OTP_i = h(OTP_i^{'})$, representing the verification of the last half factor of 2.5 factors.

	\item Service provider verifies the transaction and its inclusion proof $\pi_{inc}$.

	\item The service provider grants access to the user a and marks the Session ID as invalidated.
\end{enumerate}

\subsection{Attempt of Credential Misuse \& Reinitialization}
\label{sec:reinicialization}
During the authentication, the service provider and the user can detect an attempt of credentials misuse.
Note that authentication cannot be finished by the attacker since she does not possess precursors $OTPs^{'}$.
In such a case, the user aborts the authentication process and has to generate new credentials since the attacker already has access to the client interface of the user, from which she stole $SK_U$ and $OTPs$.
Therefore, the user first ensures that her client or PC does not contain any malware, and then repeats the bootstrapping protocol (see \autoref{fig:bp}), which will introduce a new private key and OTPs including their precursors, while the public data linked to these secrets will be updated at the service provider as well.

\section {Implementation}
\label{sec:implementation}
We made a proof-of-concept implementation of smart contract registry in Solidity and tested it on EVM-compatible blockchain while other components of our approach were programmed in Javascript.
We were mostly interested in evaluating the operational costs (in gas units) of our solution stemming from the execution of transactions on a  blockchain.

In detail, the cost of deploying   smart contract by a service provider is equal to \textbf{292k} of gas and the cost writing the OTP to the blockchain is equal to \textbf{48k} of gas.

\subsection{Performance at Different Blockchain Types}
\paragraph{\textbf{Ethereum}}
If we were hypothetically assuming a public EVM-compatible blockchain like Ethereum, the theoretical estimation of maximum possible authentication attempts would depend of the block gas limit and average block creation time, assuming that block would contain transactions only from our authentication. 
The current Ethereum's block gas limit is equal to 15M and can be dynamically extended up to 30M, while it creates a block every 12s on average.
Therefore, Ethereum would handle maximally $\frac{30M}{48k*12s} = 52$ authentication attempts per second with our solution.
However, this would be very expensive and unrealistic, since Ethereum is used by many other users.

\paragraph{\textbf{Polygon Matic}}
Therefore, we argue that our solution would be more viable in the case of second layer blockchains (such as Polygon Matic) or public permissioned blockchains (such as Hyperledger Besu).
Note that Polygon Matic creates a block every 2 second while it has 20M of block gas limit, which together implies that Polygon would maximally process $\frac{15M}{48k * 2s} = 126$ authentication attempts per second with our solution.

\paragraph{\textbf{Hyperledger Besu}}
The performance of Hyperledger Besu with QBFT was studied in~\cite{fan2022performance}, where the authors reported the throughput of around 600 smart contract invocations per second on the benchmarking simple contract and its method \textit{open},\footnote{\url{https://github.com/hyperledger/caliper-benchmarks/blob/main/src/ethereum/simple/simple.sol}} which costs 45k of gas.
This means that, with our solution, the Hyperledger Besu would be able to accommodate for 562 authentication attempts.

\section {Security Analysis}
\label{sec:security}
In this section, we make a closer look on the security of the proposed approach.
We will discuss a few potentially malicious exclusively occurring attackers, while we state the ways of how our approach protects against them or justify the reasons why it cannot provide a protection.

\subsubsection*{\textbf{Stealing the Private Key and OTPs from the Client}}
First, we consider our main attacker model, in which the attacker can steal the secrets stored at the client interface of the user -- in detail, her private key and OTPs.
The attacker might initiate the authentication session but will not be able to finish it since she does not have precursors of OTPs.
Moreover, upon malicious session initialization attempt, the data in blockchain will be updated; therefore, the user can easily detect such a situation and reinitialize the protocol as described in \autoref{sec:reinicialization}.
Note that beside the user, also the service provider can detect this situation from the blockchain and notify the user.
This shows that our approach is capable of detecting and preventing the attacker who steals the secrets from the client. 

\subsubsection*{\textbf{Stolen Authenticator}}
It is straightforward to see that this attack could not be performed on our approach.
If the attacker were to steel the user's authenticator device, she would have no way of starting the operational protocol since she does not posses the private key and OTPs of the user.

\subsubsection*{\textbf{Session Hijacking}}
A session hijacking attack is the misuse of the user's session with the service provider's site to steal their essential information.
This attack can commonly occur between a web browser and server by exploiting HTTPS session cookies.
In any case, if the attacker manages to eavesdrop a user's session, she will not see $OTP_i$ because it is encrypted by TLS, so she cannot cause any damage.
Note that this is a common practice and does not relate to our approach.

\subsubsection*{\textbf{Man-in-the-Middle Attack}}
A Man-in-the-Middle (\textit{MitM}) attack is a general term where an attacker eavesdrops or impersonates one of the parties by placing themselves between two entities, i.e., a website/application and a user, to steal personal information (e.g., login credentials).
Communication pretends to be a routine exchange of information, but all communication between the two only happens through the attacker.
We assume two different types of MitM attacks possible in our approach:
\begin{itemize}
	\item \textbf{MitM between the client and the service provider.}
		We consider that the attacker pretends that she is the service provider, luring the user to provide authentication factors.
		However, this is not possible due to verification of the service provider by the client interface/user who checks a public key of the provider from X.509 infrastructure.
		This is the standard approach in securing the client/server communication and our approach does not differ from others.
	\item \textbf{MitM between the service provider and the blockchain.}
		This kind of MitM attacker might temporarily postpone the transaction sent to the blockchain; however, due to censorship resistance nature of the blockchain (i.e., p2p gossiping network), the transaction will be eventually processed.
		If we were to assume that the attacker can fully eclipse the victim (i.e., service provider), then we would view this situation as DoS attack on the service provider, which is a common attack on any service provider and is not affected by our approach.

\end{itemize}

\subsubsection*{\textbf{Malware in the Client Interface}}
The strongest attacker that might potentially occur in our approach is the malware in the victim's PC (i.e., the client interface), which is also known as man-in-the-machine attack (MitMa)~\cite{bui2018man}. 
This attack enables tampering with the client, which entails obtaining the user secrets stored at the client as well as displaying arbitrary (spoofing) information to the user while acting in the attacker's desired way (including extraction of the current OTP precursor).
Nevertheless, we note that it is common that authentication methods cannot protect from this type of attacker model and we state it only for completeness. 
The effective protection measures involve the antivirus in the user machine and the browser.

Also note that to make this attack successful, the attacker requires the user to provide OTP from the authenticator. 
Therefore, this can be mounted only interactively along the user authentication is in progress, which has some other potential implications.
If the attacker would finish operational protocol of our approach while the benign user's authentication was in progress, a freshly authenticated user session would be dropped, which can be noticed by the user. 
Therefore, the user can observe unusual behavior and try to resolve by a new authentication attempt.
ALike the user, also service provider can observe an unusual behavior, and it can inform the user about it by other means (e.g., email, SMS, etc). 
Eventually the user might reinitialize our approach on a cleaned PC.

\section{Discussion}
\label{sec:discussion}

\subsection{Comparison with the Centralized Credential Misuse Detection}
One might argue that centralized solutions such as notification emails might serve the purpose and no blockchain is necessary. 
However, in contrast to our solution, the centralized solution cannot reliably distinguish between the malicious and benign credential misuse.
This would translate into intrusive frequent emails confirming that some authentication attempt was made, which will not serve the purpose since the user will soon or later start ignoring them.
If a centralized solution would filter the number of emails by assuming a change in IP addresses or location, the number of emails might be decreased but user changing the IP addresses (or machines/devices) frequently might fallback into into ignoring even such emails, while in contrast our solution would have guaranteed maximal true positive rate.
Another issue that is of concern to notification emails is possibility of phishing attacks.

\subsection{Costs and Performance of the Blockchain}
Since our solution utilize smart contract platform for its execution, service provider has to pay for the deployment and execution of the smart contract.
These costs might be too high in the case of using the public permissionless blockchain, such as Ethereum.
Therefore, we recommend to use a public permissioned blockchains such as Hyperledger Besu (as we suggested in \autoref{sec:implementation}) that might be operated by various service provides utilizing our solution.

On the other hand, multiple service providers using the same blockchain might cause decrease in the number of maximum authentication attempts per second. 
However, we argue that the user sessions should have a long enough time for their expiration, causing the new authentication attempt to occur only once in a day (or a few days). 
This would not hurt the capability of detecting credential misuse, which would still remain immediate.

\subsection{Storage Overhead of the Blockchain}
Authentication is very frequent operation and assuming that some services might have millions of users, the storage of the blockchain is important to be optimized.
Therefore, we keep only the last used OTP for each user at the blockchain.
Assuming the air-gapped setting of SmartOTPs that work with 16B long OTPs, for example in the case of one million users, the storage overhead in the state tree of the smart contract platform would be only $\sim$16MB.

On the other hand, one has to account also for the storage overhead of the transactions related to authentication, where for each authentication attempt a transaction is created.
However, in contrast to state tree of the smart contract platform, the storage for transaction can always be slower and thus cheaper, which makes our approach applicable (especially in the case of consortia permissioned blockchain, as discussed above).

\subsection{Account Recovery}
For simplicity of description, we assumed in the bootstrapping phase of our approach (see \autoref{sec:bootstrapping}) that the user creates a new public/private key par which is register with identity provider. 
However, this would cause that to reinitialize our approach, the user would have to register the new keypair with the identity provider, which might take some time and cost some amount of money.
Therefore, to optimize this approach, we propose to bootstrap our scheme with the new keypair generated by the user, which is signed by the original keypair registered at service provider. 
In this way, the user will be able to reinitialize our approach without the need of contacting the identity provider.

\subsection{Usability Implications}
Our approach imposes usability consequences similar to other two factor authentication approaches such as Google Authentication or SmartOTPs.
Hence, the user requires her smart phone (or other device storing OTP precursors $OTPs'$).
On the other hand the usability related to transferring of OTPs is equivalent to SmartOTPs, and we refer the reader to the paper introducing this approach~\cite{b13}.

\section{Related Work}
\label{sec:related}
There exist many authentication schemes for centralized service providers and in this section, we briefly review them.
Then, in the second part of this section we briefly review blockchain-based identity management solutions.

\subsection{Authentication Schemes}
We daily use several authentication methods that are based on a single factor, such as username/password, PIN number, delegated authentication via OAuth~\cite{OAuth}, smart-card authentication~\cite{smartcard2}, biometric characteristics\footnote{Login to mobile or laptop operating system.} (e.g., fingerprint, voice, iris, retina)~\cite{bio1, bio2}, personal access tokens (i.e., GitHub)~\cite{github}, fingerprinting of the computer\footnote{Including information such as IP address, location, computer model, and browser cookies.}~\cite{gpatent1, gpatent2}, etc.
We refer the reader to the work of Malinka et al.~\cite{banking} for an overview of several authentication primitives in a banking environment.

\subsubsection{\textbf{Two Factor Authentication (2FA)}}
A common method to secure authentication usually involve a combination of multiple factors. 
These combinations of authentication factors might be, for example, as follows:
\begin{itemize}
	\item Email/Password \& SMS
	\item Password \& HW Token
	\item Email/SMS \& Application confirmation
	\item Smart-card-based password~\cite{smart_card}
	\item HW Token \& Email/SMS \& Application (3 factor)
\end{itemize}

\medskip\noindent
The two most common multi-factor authentication protocols~\cite{10.1145/3440712} used by users are Google's 2-Step authentication, Google Authenticator with OTPs, and FIDO's U2F.
In addition to these solutions, there are various implementations such as Authy~\cite{authy} or LastPass Authenticator~\cite{lastpass}.
We briefly describe some of them in the following.

\subsubsection*{\textbf{Google 2-Step}}
Google proposes a two-factor authentication mechanism called \textit{Google 2-Step}~\cite{google}.
The user uses the \textbf{phone} to confirm login (identity verification).
\textit{Google 2-Step} has several variations.
The default mechanism is that an authentication code is sent to the user via SMS, Google 2-Step with \textit{Verification Codes}.
Then the received code needs to be entered into the user's computer.
An alternative is the Google 2-Step with \textit{One-Tap} version, where the user presses the \textit{Yes} button in the application pop-up.
The server creates a fresh random token which is sent to the user's mobile application.
Difference is that communication between the server and the phone is over a TLS channel rather than by SMS.

\subsubsection*{\textbf{FIDO's U2F}}
FIDO is an alliance that aims to provide standards for secure authentication.
It proposes many solutions under the \textit{FIDO}, \textit{FIDO2} and \textit{U2F} standards (a.k.a., WebAuthn)~\cite{fido1}.
One proposed solution is the \textit{Universal 2nd Factor (U2F)} protocol, that enables relying parties to offer a strong cryptographic 2nd factor option for end user security.
The relying party's dependence on passwords is reduced.
For example the password can be a simplified PIN digit.
The FIDO U2F also offers a variation with OTP generated HW token.

\subsubsection*{\textbf{OTPs (One Time Passwords)}}
The 2FA methods also include the use of OTPs.
Probably the best known product in this domain is the \textit{Google Authenticator}~\cite{gauth} protocol.
In this protocol, the phone and the server share a secret key and use it to derive an OTP token.
Depends on the variation the token could be derived from the \textbf{current time} or based on \textbf{counter}.
In other words Google Authenticator is based on \textit{Time-Based OTPs}~\cite{totp} (TOTP) or \textit{HMAC-based OTPs}~\cite{hotp} (HOTP, hash chains) for authenticating users.
\textit{S/KEY}~\cite{skey} OTP system for Unix-like operating systems is based on the same principle.
In terms of hardware implementation, the \textit{FIDO U2F} variant mainly focuses on the use of the USB token as the second factor.
The U2F protocol relies on a device capable of securely generating and storing secret and public keys and performing cryptographic operations using these keys.
For example, one implementation of FIDO U2F is the Yubico YubiKey \textbf{hardware device}~\cite{yubico}, which generates cryptographically signed tokens.
The hardware device has a button that the user must press to confirm the transaction.
To enable second factor authentication for some service, the device generates a key pair, and the public key is registered within the service server (the service and the device are binded).

\subsection{Blockchain-Based Identity Management}
Blockchain can be used to manage the verifiable credentials and to be the future base of modern solutions of identity management.

\subsubsection*{\textbf{Sovrin}}
Sovrin~\cite{sovrin} is a trust framework for decentralized, global public solution, based on self-sovereign identity management.
Sovrin provides portable identity for all its users. Verifiable credentials contain some private information, therefore they are disclosed only to service providers that request them.
They might contain information such as  gender, education, age, employment history, etc.
The Sovrin is based on Hyperledger Indy~\cite{hindy} blockchain, which represent an example of private permissioned blockchain.
Hyperledger Ursa is used for cryptographic operations, and Hyperledger Aries as agent that stores verifiable credentials.
The Sovrin uses decentralized identifiers (DIDs) as identificators of relationships between users and service providers.

\subsubsection*{\textbf{uPort}}
uPort~\cite{b14} is a system based on Ethereum blockchain and adopts the principles of self-sovereign identity.
The essence of the uPort identity is the Ethereum account address on which users interact, and the identity is permanent.
Users can send and request credentials, digitally sign transactions and manage identity.
It enables to publish identity data on another blockchain.

\subsubsection*{\textbf{ShoCard}}
ShoCard (presently known as PingOne)~\cite{pingone} is a blockchain-based identity magement system where users can store and protect their digital identities.
The user's identity information will always be used together (bound) with the user's key to ensure privacy.
This eliminates the need for a third-party database.
ShoCard stores the authentication code of the user's data in the blockchain, which can guarantee the legitimacy of the personal identity and facilitate third-party authentication.

\subsubsection*{\textbf{MyData}}
Mydata~\cite{b15} is a project of Finnish government for managing personal data. The project is based on MyData authentication and user managed access, OpenID single sign-on and Oauth 2.0 which control access to Web APIs.
Alike Sovrin, MyData uses Hyperledger Indy Blockchain.

\subsubsection*{\textbf{I/O Digital}}
I/O Digital (I/O Coin)~\cite{iocoin} is a blockchain-based identity management framework, which uses its own DIONS
(Decentralized I/O Name Server) and Proof of Stake (PoS I/O). The DIJONS blockchain, and its implementation IOC, allows storage of data up to 1MB size. DIJONS uses AES-256 block cipher and it has a feature called Alias system. Alias system is used for managing reputation and control data by users. The project will migrate to Ethereum-compatible blockchain type.

\subsubsection*{\textbf{Other Solutions Based on Ethereum}}
The relevant identity management solutions based on Ethereum are Waypoint~\cite{b18}
BlockStack~\cite{b19}, UniquID~\cite{b23}, Jolocom~\cite{b24} and Authenteq~\cite{b27}.
There are several works (surveys) by Kuperberg~\cite{surv3}, Liu et al.~\cite{surv2}, Xiaoyang et al.~\cite{surv1} that attempt to comprehend multiple different blockchain-based identity management schemes and frameworks.

\section{Conclusion}
\label{sec:conclusion}
Identity management based on blockchain technology offers modern solution to access and manage data with respect to security and privacy of users.
We assume such systems used by centralized service providers for various reasons, while one of them is credentials-based authentication with private keys.
We propose an approach that enriches such an authentication by 1.5 factors obtained from state-of-the-art approach called SmartOTPs.
We slightly modify SmartOTPs to contain only 2 hash chain layers, where the first layer represents precursors of OTPs and the second layer is derived from the first one by cryptographically secure hashing and represents the OTPs themselves.
To perform authentication, OTPs are provided as the second factor, while their precursors are in the later message provided as the half factor that finishes the authentication process.
We demonstrated that with our approach, the user can detect stoled and misused credentials from her client immediately.
Our approach is applicable in the setting of any centralized service provider and to make it cheap and efficient, we recommend the use a public permissioned blockchain running by multiple service providers that want to benefit from our solution.

\begin{acks}
  The authors would also like to thank the anonymous referees for
  their valuable comments and helpful suggestions. The work is
  supported by the \grantsponsor{GS501100001809}{National Natural
    Science Foundation of
    China}{http://dx.doi.org/10.13039/501100001809} under Grant
  No.:~\grantnum{GS501100001809}{61273304}
  and~\grantnum[http://www.nnsf.cn/youngscientsts]{GS501100001809}{Young
    Scientsts' Support Program}.
\end{acks}
 
\bibliographystyle{ACM-Reference-Format}
\bibliography{bibliography}

\end{document}